\documentclass[aps,twocolumn,pra,floatfix]{revtex4-2}
\usepackage{epsfig}
\usepackage{graphicx}
\usepackage{dcolumn}
\usepackage{amssymb,amsmath}
\usepackage{mathrsfs}

\usepackage{bm,amsfonts,mathtools}

\usepackage{tikz}
\usepackage{xcolor}

\begin{document}

	\title{Calculation of the hyperfine structure of Dy, Ho, Cf, and Es}
	
	\author{Saleh O. Allehabi, V. A. Dzuba, and V. V. Flambaum}
	
	\affiliation{School of Physics, University of New South Wales, Sydney 2052, Australia}

	\begin{abstract}
		 A recently developed version of the configuration interaction (CI) method for open shells with a large number of valence electrons has been used to study two heavy atoms, californium (Cf, Z= 98)  and  einsteinium (Es, Z= 99). 
Motivated by experimental work to measure the hyperfine structure (HFS) for these atoms, we perform the calculations of the magnetic dipole HFS constants $A$ and electric quadrupole HFS constant $B$ for the sake of interpretation of the measurements in terms of nuclear magnetic moment $\mu$ and electric quadrupole moment $Q$. 
For verification of our computations, we have also carried out similar calculations for the lighter homologs dysprosium (Dy, Z= 66) and holmium (Ho, Z= 67), whose electronic structures are similar to Cf and Es, respectively. We have conducted a revision of the nuclear moments of some isotopes of Es leading to an improved value of the magnetic moment of $^{253}$Es [$\mu$($^{253}$Es) = 4.20(13)$\mu_N$].
		
	\end{abstract}
		\date{\today}
	
	\maketitle

	\section{Introduction}

The study of atomic properties of heavy actinides has gained growing interest ~\cite{Es1975,Es2009,Raeder2022,Mustapha2022,Fm1,Fm2,ErFm,Raeder}. 
Transition frequencies and hyperfine structure (HFS) are being measured. Measuring HFS is motivated by obtaining data on the nuclear momenta of heavy nuclei. This would advance our knowledge about the nuclear structure of superheavy nuclei benefiting the search for the hypothetical stability island.
In light of this, we focus on theoretically studying of the hyperfine structure for heavy actinides, californium (Cf, Z= 98) and einsteinium (Es, Z= 99). Combining the calculations with the measurements would allow the extraction of the nuclear magnetic moment $\mu$ and electric quadrupole moments $Q$ of the studied isotopes.

HFS constants of some states of odd isotopes of Cf ($^{249}$Cf,$^{251}$Cf,$^{253}$Cf) were recently measured and nuclear moments $\mu$ and $Q$ were extracted using our calculations~\cite{Raeder}. This work presents a detailed account of these calculations as well as similar calculations for Es.
In the case of  Es, there are no theoretical results currently available, whereas several experimental papers have been published. Using different empirical techniques, Refs.~\cite{Es1975,Es2009,Raeder2022} studied the HFS of Es for three isotopes with non-zero nuclear spins, $ ^{253,254,255}$Es.

Heavy actinides like Cf and Es, are atoms with an open $5f$ subshell. The number of electrons on open shells is twelve for Cf and thirteen for Es (including the $7s$ electrons).This presents a challenge for the calculations. We use the configuration interaction with perturbation theory (CIPT)~\cite{cipt} method, which has been developed for such systems.  To check the applicability of the method and the expected accuracy of the results we performed similar calculations for lanthanides dysprosium (Dy, Z= 66) and holmium (Ho, Z= 67), whose electronic structures are similar to Cf and Es, respectively. 
Both, Dy and Ho were extensively studied experimentally and theoretically (see, e.g. \cite{Dy1970,NIST,Dy1967,Dy1974,DzubaDy,Cheng85,Ho,hfs_Ho,DzubaHo}).
Here we compare our results to experimental data, Refs.~\cite{Dy1970,NIST} for Dy and Refs.~\cite{NIST,Ho,hfs_Ho} for Ho, to check the accuracy of the method we use.

	\section{Method of calculation}
	
	\subsection{Calculation of energy levels}
	
	As it was mentioned in the introduction the Dy and Cf atoms have twelve valence electrons, the Ho and Es atoms have thirteen valence electrons. It is well known that as the number of valence electrons increases, the size of the CI matrix increases dramatically, making the standard CI calculations practically impossible for such systems. In this work we use the CIPT method~\cite{cipt} which has been especially developed for such systems. It reduces the size of the CI matrix by neglecting the off-diagonal matrix elements between high-energy states and reducing the contribution of these states to the perturbation theory-like corrections to the matrix elements between low-energy states. The size of the resulting CI matrix is equal to the number of low-energy states.
	
The CI Hamiltonian can be written as follows
\begin{equation}
	\hat{H}^{\mathrm{CI}}=\sum_{i=1}^{N_{v}}\hat{H}^{\mathrm{HF}}_{i}+\sum_{i<j}^{N_{v}}\frac{e^{2}}{\left|r_{i}-r_{j}\right|},
\end{equation}
where $ i $ and $ j $ enumerate valence electrons and  $ N_{v} $ is the total number of valence electrons, $ e $ is electron charge, and $ r $ is the distance. $ \hat{H}^{\mathrm{HF}}_{i} $ is the single-electron Hartree-Fock (HF) Hamiltonian, which has the form 
\begin{equation} \label{e:RHF}
	\hat{H}^{\mathrm{HF}}_{i}= c{ \bm{\alpha}}_i\cdot {\bf \hat p}_i+(\beta -1)mc^2+V_{\rm nuc}({r_i})+V^{N-1}({r_i}).
\end{equation}
Here $c$ is the speed of light, $\bm{\alpha}_i$ and $\beta$ are the Dirac matrixes, $\bf \hat p_i$ is the electron momentum, $m$ is the electron mass, $V_{\rm nuc}({i})$ is the nuclear potential obtained by integrating Fermi distribution of nuclear charge density, and $V^{N-1}({i})$  is the self-consistent HF potential obtained for the configuration with one $7s$ (or $6s$) electron removed from the ground state configuration of the considered atom. This corresponds to the $V^{N-1}$ approximation~\cite{VN1,VN2} which is convenient for generating a single-electron basis. Single-electron basis states are calculated in the frozen $V^{N-1}$ potential, so that they correspond to the atom with one electron excited from the ground state.
External electron wave functions are expressed in terms of coefficients of expansion over single-determinant basis state functions 
\begin{eqnarray}
&&\Psi (r_1, \dots ,r_M)=   \label{e:Psi}  \\
&&\sum_{i=1}^{N_1} X_i \Phi_i (r_1, \dots ,r_M)+ \sum_{j=1}^{N_2} Y_j \Phi_j(r_1, \dots ,r_M). \nonumber
\end{eqnarray}
Here $M$ is the number of valence electrons, $N_1$ is the number of low-energy basis states, $N_2$ is the number of high-energy basis states.

Then the CI matrix equation can be written in a block form
\begin{equation} \label{e:blocks}
\left( \begin{array}{cc} A & B \\ C & D \end{array} \right) \left(\begin{array}{c} X \\ Y \end{array} \right) = E_a \left(\begin{array}{c} X \\ Y \end{array} \right).
\end{equation}
Here block $A$ corresponds to low-energy states, block $D$ corresponds to high-energy states, and~blocks $B$ and $C$ correspond to cross terms. 
Note that since the total CI matrix is symmetric, we have $C = B'$, i.e.,~$c_{ij} = b_{ji}$. 
Vectors $X$ and $Y$ contain the coefficients of expansion of the valence wave function over the single-determinant many-electron basis functions (see Eq.~\ref{e:Psi}).

Finding $Y$ from the second equation of (\ref{e:blocks}) leads to
\begin{equation}\label{e:Y}
Y=(E_aI-D)^{-1}CX.
\end{equation}
Substituting $Y$ to the first equation of (\ref{e:blocks}) leads to
\begin{equation}\label{e:CIPT}
\left[A + B(E_aI-D)^{-1}C\right] X = E_a X,
\end{equation}
where $I$ is  the unit matrix. 
Then, following Ref.~\cite{cipt} we neglect off-diagonal matrix elements in block $D$. This leads to a very simple structure of the $(E_aI-D)^{-1}$ matrix, $(E_aI-D)^{-1}_{ik} = \delta_{ik}/(E_a - E_k)$, where $E_k = \langle k|H^{\rm CI} |k \rangle$.
Matrix elements of the effective CI matrix (\ref{e:CIPT}) have the form
\begin{equation}
	\langle i|\hat H^\mathrm{eff}|j\rangle =\langle i|\hat H^\mathrm{CI}|j\rangle + \sum_{k}\frac{\langle i|\hat H^\mathrm{CI}|k\rangle\langle k|\hat H^\mathrm{CI}|j\rangle}{E_a-E_{k}}.
	\label{e:CIPT2}
\end{equation}
We see that the standard CI matrix elements between low-energy states are corrected by an expression which is very similar to the second-order perturbation theory correction to the energy. This justifies the name of the method. To calculate this second-order correction we need to know the energy of the state $E_a$ which must come as the result of the solution of the equation, i.e. it is not known in advance. Therefore, iterations are needed. We start from any reasonable guess for the energy. For example, it may come from the solution of the equation with neglected second-order correction. Note that the energy independent numerators of the second-order correction can be calculated only once, on the first iteration, kept on disk and reused on every consequent iteration. This means that only the first iteration takes some time while all other iterations are very fast. As a rule, less than ten iterations are needed for full convergence. As a result, we have an energy of the state $E_a$ and expansion coefficients $X$ and $Y$.

\subsection{Basis states}

To solve the CI equations we need many-electron basis states which are constructed from single-electron states.
For single-electron basis states we use the B-spline technique~\cite{Johnson_Bspline,Johnson_Bspline2}.
These states are defined as linear combinations of B-splines that are eigenstates of the HF Hamiltonian (\ref{e:RHF}). 
Forty B-splines of the order nine are calculated within a box of radius  $R_{\rm max}$ = 40$a_B$ (where $a_B$ represents Bohr's radius) and an orbital angular momentum of 0~$\leq$~\textit{l}~$\leq$~4. 14 states above the core in each partial wave  are used. It has been found that by selecting the values of $l_{\rm max}$, $R_{\rm max}$, and the number of B-splines, a basis is adequately saturated for low-lying states. The many-electron states are found by making all possible single and double electron excitations from a few reference configurations. One, two or three configurations, corresponding to the low-lying states of an atom are considered as reference configurations. One configuration of the same parity is considered at a time.
For each configuration, all possible values of the projection of the total angular momentum $j$ of the single-electron states are considered and many-electron states with fixed values of total many-electron angular momentum $J$ and its projection $M$ are constructed. Usually, we take $M=J$.

\subsection{Calculation of hyperfine structure}

In this section, we mostly follow our previous work on hafnium and rutherfordium~\cite{HfRf}.
To calculate HFS,
 we use the time-dependent Hartree-Fock (TDHF) method, which is equivalent to the well-known random-phase approximation (RPA).
The RPA equations are the following:	
\begin{equation}\label{e:RPA}
	\left(\hat H^{\rm RHF}-\epsilon_c\right)\delta\psi_c=-\left(\hat f+\delta V^{f}_{\rm core}\right)\psi_c
\end{equation}
where $\hat f$ is an operator of  an external field (nuclear magnetic dipole or electric quadrupole fields).  
Index $c$ in (\ref{e:RPA}) numerates states in the core, $\psi_c$ is a single-electron wave function of the state $c$ in the core, $\delta\psi_c$ is the correction to this wave function caused by an external field, and $\delta V^{f}_{\rm core}$ is the correction to the self-consistent RHF potential caused by changing of all core states. 
Eq. (\ref{e:RPA}) are solved self-consistently for all states in the core. As a result, an effective operator of the interaction of valence electrons with an external field is constructed as $\hat f + \delta V^{f}_{\rm core}$. The energy shift of a many-electron state $a$ is given by
\begin{equation} \label{e:de}
\delta \epsilon_a = \langle a | \sum_{i=1}^M \left(\hat f+\delta V^f_{\rm core} \right)_i | a\rangle.
\end{equation}
Here $M$ is the number of valence electrons.

When the wave function for the valence electrons comes as a solution of Eq.~(\ref{e:CIPT}), Eq.~(\ref{e:de}) is reduced to
\begin{equation}\label{e:mex}
\delta \epsilon_a = \sum_{ij} x_i x_j \langle \Phi_i|\hat H^{\rm hfs}|\Phi_j \rangle,
\end{equation}
where $\hat H^{\rm hfs} =  \sum_{i=1}^M (\hat f+\delta V^f_{\rm core})_i$.
For better accuracy of the results, the full expansion (\ref{e:Psi}) might be used. Then it is convenient to introduce  a new vector $Z$, which contains both $X$ and $Y$, $Z \equiv \{X,Y\}$. Note that the solution of (\ref{e:CIPT}) is normalized by the condition $\sum_i x_i^2=1$. The normalization condition for the total wave function (\ref{e:Psi}) is different,  $\sum_i x_i^2+\sum_j y_j^2 \equiv \sum_i z_i^2=1$. Therefore, when $X$ is found from (\ref{e:CIPT}), and $Y$ is found from (\ref{e:Y}), both vectors should be renormalized. Then the HFS matrix element is given by the expression, which is similar to (\ref{e:mex}) but has much more terms 
\begin{equation}\label{e:mez}
\delta \epsilon_a = \sum_{ij} z_i z_j \langle \Phi_i|\hat H^{\rm hfs}|\Phi_j \rangle.
\end{equation}

Energy shift (\ref{e:de}) is used to calculate HFS constants $A$ and $B$ using textbook formulas
\begin{equation}
A_a = \frac{g_I \delta \epsilon_a^{(A)}}{\sqrt{J_a(J_a+1)(2J_a+1)}},
\label{e:Ahfs}
\end{equation}
and
\begin{equation}
B_a = -2Q \delta \epsilon_a^{(B)}\sqrt{\frac{J_a(2J_a-1)}{(2J_a+3)(2J_a+1)(J_a+1)}}. 
\label{e:Bhfs}
\end{equation}
Here $\delta \epsilon_a^{(A)}$ is the energy shift (\ref{e:de})  caused by the interaction of atomic electrons with the nuclear magnetic moment $\mu$, $g_I=\mu/I$, $I$ is nuclear spin; $\delta \epsilon_a^{(B)}$ is the energy shift (\ref{e:de}) caused by the interaction of atomic electrons with the nuclear electric quadrupole moment $Q$ ($Q$ in (\ref{e:Bhfs}) is measured in barns). 

\section{Energy levels and HFS of Dysprosium  and Holmium}
	
		\begin{table}
		
		\caption{\label{t:Er}
			Excitation energies ($E$, cm$^{-1}$), and $g$-factors for some low states of Dy, and Ho atoms.} 
		\begin{ruledtabular}
			\begin{tabular}{c cc cc cc }
				&&&
				\multicolumn{2}{c}{This work } &
				\multicolumn{2}{c}{NIST \cite{NIST}}\\
				\cline{4-5}
				\cline{6-7}
				
				\multicolumn{1}{c}{Conf.}&
				\multicolumn{1}{c}{Term}&
				\multicolumn{1}{c}{J}&
				\multicolumn{1}{c}{$ E $}&
				\multicolumn{1}{c}{$ g $}&
				\multicolumn{1}{c}{$ E $}&
				\multicolumn{1}{c}{$ g $}\\
				\hline
				
					\multicolumn{7}{c}{\textbf{Dy}}\\
				
				$4f^{10}6s^2$&$ ^{5}$I &8& 	  0.000 &1.242&   0.000  & 	  	  1.2416\\
				$4f^{10}6s^2$&&7&3933& 1.175&	  4134.2 & 1.1735\\
				$4f^{10}6s^2$&& 6 	       	  	  &7179& 1.073&	  7050.6 & 1.0716\\
				$4f^{9}5d6s^2$& $ ^{7} $H$ ^{o} $&8 &7818&1.347 & 	  7565.610&  1.35246 	\\
				$4f^{9}5d6s^2$& &7 &9474 &  1.353 & 	  8519.210&1.336\\
				$4f^{10}6s^2$&$ ^{5}$I&5 	        	  	  &9589 & 0.909&	   9211.6 & 0.911\\
				$4f^{9}5d6s^2$& $ ^{7} $I$ ^{o} $&9 &10048&  1.316 & 	  9990.974       	  	&  1.32\\
				$4f^{9}5d6s^2$& $ ^{7} $H$ ^{o} $&6 &11052&1.417 & 	  10088.802 &      	  	  1.36\\
				$4f^{10}6s^2$&$ ^{5}$I&4&11299 & 0.613&	   	  10925.3       	  	 & 0.618\\

				\hline
				
					\multicolumn{7}{c}{\textbf{Ho}}\\
			
				$4f^{11}6s^2$& $ ^{4} $I$ ^{o} $& 	  15/2 	 &0.00&1.196&0.00&1.1951\\	 
				$4f^{11}6s^2$& & 	   13/2 	  &5205  & 1.107&  5419.7  &$ - $\\	 
				
				 $4f^{10}5d6s^2$&	 (8,$\frac{3}{2}$) &	 17/2&8344&1.262& 	  8378.91&$ - $\\
				$4f^{10}5d6s^2$& &15/2 &8385	&1.280  &8427.11&$ - $\\
								$4f^{11}6s^2$& $ ^{4} $I$ ^{o} $& 	   11/2 	  &8501  &0.979&   8605.2 & 1.012  \\	 
				$4f^{10}5d6s^2$& (8,$\frac{3}{2}$)&13/2 	 &	8989&1.336  &9147.08&$ - $\\
					$4f^{10}5d6s^2$& &19/2 &8952	& 1.231 &9741.50&$ - $\\
				
				$4f^{11}6s^2$& $ ^{4} $I$ ^{o} $& 	   9/2 	  &10550  & 0.780&   10695.8 &0.866\\	 
				
			\end{tabular}
		\end{ruledtabular}
	\end{table}

	\begin{table*}
		\caption{\label{t:HFS}
			Hyperfine structure constants $A$ and $B$ (in MHz) for low-lying states of Dy and Ho. Nuclear spin $I$, nuclear magnetic moment $\mu(\mu_N)$, and
			nuclear electric quadrupole moment $Q(b)$ values for the isotopes of the $ ^{161,163} $Dy and $ ^{165} $Ho are taken from Ref.~\cite{Stone1}, 
			$g_I=\mu/I$. Last column presents references to experimental data for $A$ and $B$.} 
		\begin{ruledtabular}
			\begin{tabular}{cc  cc c cc c c}

				\multicolumn{1}{c}{Isotope}&&&&
				\multicolumn{2}{c}{ This work}&
				
				\multicolumn{3}{c}{Experimental results.}\\
				\cline{5-6}
				\cline{7-9}
				\multicolumn{1}{c}{Nuclear Parameters}&
				\multicolumn{1}{c}{Conf.}&
				\multicolumn{1}{c}{Term}&
				\multicolumn{1}{c}{$ J $}&
				
				\multicolumn{1}{c}{$A$}&
				\multicolumn{1}{c}{$B$}&

				\multicolumn{1}{c}{$A$}&
				\multicolumn{1}{c}{$B$}&
				\multicolumn{1}{c}{Ref.}\\
				
				\hline

			\multicolumn{1}{c}	{\bf $ ^{161} $Dy}  \\     

				$\mu$= -0.480, I= 5/2, Q= 2.51&$4f^{10}6s^2$&$ ^{5}$I &8 &-113&1127  & 	  -116.231	  &1091.577 &\cite{Dy1970}\\
				&$4f^{10}6s^2$&&7& -125&1057&-126.787  &1009.742&\cite{Dy1970}\\
				&	$4f^{10}6s^2$&& 6 &-140 &	991 &-139.635 &960.889&\cite{Dy1970}\\
				&$4f^{9}5d6s^2$& $ ^{7} $H$ ^{o} $&8 &-88&2256 & 	  -&-&-	\\
				&$4f^{9}5d6s^2$& &7 &-104&2397 & 	  -&-&-\\
				&	$4f^{10}6s^2$&$ ^{5}$I&5  &-166 &928 &-161.971 &894.027&\cite{Dy1970}\\
				&$4f^{9}5d6s^2$& $ ^{7} $I$ ^{o} $&9 &-80&2663 &- 	  &-&-\\
				&$4f^{9}5d6s^2$& $ ^{7} $H$ ^{o} $&6 &-122&2901 &- 	 &-&-\\
				&	$4f^{10}6s^2$&$ ^{5}$I&4 &-216&997&-205.340 &961.156&\cite{Dy1970}\\

				\hline
				
				\multicolumn{1}{c}	{\bf $ ^{163} $Dy} \\
				
				$\mu$= 0.673, I= 5/2, Q= 2.65	&$4f^{10}6s^2$&$ ^{5}$I &8 &158&1190  & 	 162.754 &1152.869&\cite{Dy1970}\\
				&$4f^{10}6s^2$&&7& 176&1116 &177.535 &1066.430&\cite{Dy1970}\\
				&	$4f^{10}6s^2$&& 6 & 196&	1046 & -&-&-\\
				&$4f^{9}5d6s^2$& $ ^{7} $H$ ^{o} $&8 &123& 2381& -	  &-&-	\\
				&$4f^{9}5d6s^2$& &7 &146&2531 & 	-  &-&-\\
				&	$4f^{10}6s^2$&$ ^{5}$I&5 &233 &979 &- &-&-\\
				&$4f^{9}5d6s^2$& $ ^{7} $I$ ^{o} $&9 &112& 2812& 	-  &-&-\\
				&$4f^{9}5d6s^2$& $ ^{7} $H$ ^{o} $&6 &170&3063 & 	 -&-&-\\
				&	$4f^{10}6s^2$&$ ^{5}$I&4 &303&1053       	  	 &- &-&-\\
				
				\hline
			\multicolumn{1}{c}	{\bf $ ^{165} $Ho}\\
				
				$\mu$= 4.17, I= 7/2, Q= 3.58	&	$4f^{11}6s^2$& $ ^{4} $I$ ^{o} $& 15/2 	 &787&-1943&800.583&-1668.089&\cite{Ho}\\	 
				&$4f^{11}6s^2$& & 	   13/2  & 939&-1668&937.209&-1438.065&\cite{Ho}\\

				& $4f^{10}5d6s^2$&	 (8,$\frac{3}{2}$) &	 17/2&666& 1085&776.4(4.5)&608(300) &\cite{hfs_Ho}\\
				&$4f^{10}5d6s^2$& &15/2 	&763 & 1127&783.0(4.5)&801(300)&\cite{hfs_Ho}\\
								&$4f^{11}6s^2$& $ ^{4} $I$ ^{o} $& 	   11/2  &1061&-1315 &1035.140 &-1052.556&\cite{Ho}  \\	
				&$4f^{10}5d6s^2$& (8,$\frac{3}{2}$)&13/2& 879 &1829&916.6(0.5)&2668(7)&\cite{hfs_Ho}\\
				&$4f^{10}5d6s^2$& &19/2 	&617 &1650 &745.1(1.4)&1747(78) &\cite{hfs_Ho}\\
				 
				&$4f^{11}6s^2$& $ ^{4} $I$ ^{o} $&9/2 &1279&-1174 &1137.700&-494.482&\cite{Ho}\\	 
				
			\end{tabular}			
		\end{ruledtabular}
	\end{table*}

For the purpose of testing the accuracy of the method, we start calculating the energy levels for some low-lying states of Dy and Ho. The results are shown in Table~\ref{t:Er}. 
	As can be seen, our results are consistent with the experimental results compiled in Ref.~\cite{NIST} of respective atomic systems. The difference between theoretical calculations and measurements is within a few hundred cm$ ^{-1} $. 
Calculated and experimental Land\'{e} g-factors are also presented. A comparison of Land\'{e} g-factors calculated with non-relativistic expressions is helpful for identifying state labels.
\begin{equation}\label{e:g}
	g_{NR} = 1 + \frac{J(J+1)-L(L+1)+S(S+1)}{2J(J+1)}.
\end{equation}
Total orbital momentum $L$, and total spin $S$ in (\ref{e:g}) cannot come from relativistic calculations. Instead, we choose their values from the condition that formula (\ref{e:g}) gives values very close to the calculated $g$-factors.
This allows us to link the state to the non-relativistic notation $^{2S+1}L_J$. Here, $J$ is the total angular momentum ({\bf J = L + S}).
A good agreement is also observed between current calculations and experimental $g$-factors of Dy and Ho whenever experimental data are available. In order to identify the states correctly, it is essential to take this into consideration. An exception stands out in state $4f^{11}6s^2\  ^{4}\rm I ^{o} _{9/2}$ of Ho, where the theory differs significantly from the experiment. Based on the NIST database~\cite{NIST} of Ho spectrum, we can observe that there are multiple states with the same parity and total angular momentum J, separated only by small energy intervals and dominated by different electron configurations. Due to this vigorous mixing, the calculations of the $g$-factor become unstable.

The hyperfine structures of the ground states and some low-lying states of Dy and Ho have also been calculated. 
The Dy atom has two stable isotopes, $^{161}$Dy and $^{163}$Dy, and the Ho atom has one stable isotope, $^{165}$Ho. The results of calculations and corresponding nuclear parameters are presented in Table~\ref{t:HFS}. One can see that we have good agreement between theory and experiment for magnetic dipole constant $A$ and electric quadrupole constant $B$ for most states of Dy and Ho. The difference between theory and experiment is within 3\% for the $A$ constant of Dy and Ho, within 4\% for the $B$ constants of Dy and $\sim$~20\% for the $B$ constant of Ho. A similar agreement between theory and experiment was found earlier for the HFS constants of Er~\cite{ErFm}.
Two states of Ho present an exception. These are the  $4f^{10}5d6s^2\ (8,\frac{3}{2})_{13/2}$ state, and the $4f^{11}6s^2\ ^{4}\rm I^{o}_{9/2}$ state.  Here the difference between theory and experiment for electric quadrupole HFS constant $B$ is significant. In particular, it is 138\% for the $4f^{11}6s^2\ ^{4}\rm I^{o}_{9/2}$ state. This is the same state which shows poor accuracy for the $g$-factor, which indicates that strong configuration mixing affects the HFS as well.
It should be mentioned that an earlier study performed using the MCDF method also found that this state had a low level of accuracy with a 117 \% deviation from the experimental result~\cite{Cheng85}.

Note that our investigations of testing the accuracy of using the CIPT method on the Er atomic system, which has a similar electronic structure, were previously performed~\cite{ErFm}. All the above atomic properties, energies, $g$-factors, and HFS constants $A$ and $B$ for the stable isotope with non-zero spin, $^{167} $Er, have been calculated. There has been a good agreement between measurements and our results (see Ref.~\cite{ErFm} Tables 1 and 6). In the end, we expect that the results for Cf and Es will be accurate as well.
	
\section{Ionization potentials}
	
		\begin{table*}
	\caption{\label{t:IP}
		Experimental and theoretical values of the first ionization potential IP$_1$ (in cm$ ^{-1} $).} 
	\begin{ruledtabular}
		\begin{tabular}{cl lc cc }
			&
			\multicolumn{2}{c}{  State}&
			\multicolumn{3}{c}{IP$_1$}\\
			\cline{2-3}
			\cline{4-6}
			
			\multicolumn{1}{c}{Atom}&
			\multicolumn{1}{c}{Initial}&
			\multicolumn{1}{c}{Final}&
			\multicolumn{1}{c}{Presesnt}&
			
			\multicolumn{1}{c}{Expt.}&
			
			\multicolumn{1}{c}{Ref.}\\
			\hline
			
			Dy&	$4f ^{10} 6s ^{2} $ \	 $ ^{5}\rm I_{8} $  & $4f ^{10} 6s $ \	 $ (8,\frac{1}{2})_{17/2} $&46658& 47901.76(5)&\cite{IPDy}\\
			Ho&$4f ^{11} 6s ^{2} $ \	 $ ^{4}\rm I^{o}_{{15}/{2}} $& $4f ^{11} 6s  $ \	 $ (\frac{15}{2},\frac{1}{2})^{\rm o}_{8} $&47819&48567(5)&\cite{IPHo} \\
			Cf&$5f ^{10} 7s ^{2} $ \	 $ ^{5}\rm I_{8} $  & $5f ^{10} 7s $ \	 $ ^{6}\rm I_{{17}/{2}} $&50821& 50663(2)&\cite{IP} \\
			Es&$5f ^{11} 7s ^{2} $ \	 $ ^{4}\rm I^{o}_{{15}/{2}} $& $5f ^{11} 7s  $ \	 $ ^{5}\rm I^{o}_{8} $&51763&51358(2)&\cite{IP} \\
			&&&&$51364.58(14) _{\rm stat} (50)_{\rm sys}$&\cite{IPEs}\\

			
		\end{tabular}
	\end{ruledtabular}
	
\end{table*}

Calculating ionization potential (IP) is a good way to test the theoretical approach for the ground state. The IP is obtained as a difference between the ground state energies of the neutral atom and the ion. The CIPT method, which we use in the present calculations, has a feature of having good accuracy for low-lying states, and it decreases while going up on the energy scale. The best accuracy is expected for the ground state. On the other hand, having HFS for the ground state is sufficient to extract nuclear parameters $\mu$ and $Q$. Therefore, we calculate the first ionization potential (IP$_1$) for all atoms considered in the present work. We calculate ground state energies of neutral atoms and corresponding ions in the same $V^{N-1}$ potential and the same single-electron basis. 
This ensures exact cancelation of the energies associated with core electrons. 
The results are presented in Table~\ref{t:IP} and compared with available experimental data. 
As can be seen from the table the accuracy of the results is 2.7\% for Dy, 1.6\% for Ho, 0.3\% for Cf and 0.8\% for Es.
		
		
\section{Results for HFS}
	
\begin{table}
\caption{\label{t:GHFS}
		Calculated hyperfine structure constants $A$ and $B$ (in MHz) for the ground states of Dy, Ho, Cf and Es atoms.}
	\begin{ruledtabular}
		\begin{tabular}{cc  cc cc}
			
			\multicolumn{1}{c}{Atom}&
			\multicolumn{1}{c}{Conf.}&
			\multicolumn{1}{c}{Term}&
			\multicolumn{1}{c}{$ J $}&
			
			\multicolumn{1}{c}{$A$}&
			\multicolumn{1}{c}{$B$}\\
			
			\hline
			
			{ Dy} 	&	$4f^{10}6s^2$&$ ^{5}$I &8                    &587$\times$$ g _I$&449$\times$Q\\
			
			{ Ho}	& $4f^{11}6s^2$& $ ^{4} $I$ ^{o} $& 15/2 	 &661$\times$$ g _I$&-543$\times$Q\\	 

			{ Cf} 	&	$5f^{10}7s^2$&$ ^{5}$I &8                     &608$\times$$ g _I$&477$\times$Q\\
			
			{ Es}	& $5f^{11}7s^2$& $ ^{4} $I$ ^{o} $& 15/2 	 &681$\times$$ g _I$&-818$\times$Q\\	 
			
		\end{tabular}			
	\end{ruledtabular}
\end{table}

In Table~\ref{t:GHFS}, we present the results of our calculations of the HFS constants of the ground states of Dy, Ho, Cf and Es. We have calculated both, magnetic dipole HFS constant $A$ and electric quadrupole HFS  constant $B$, which can be used for the extraction of nuclear moments for any isotope with non-zero spin.
For a better understanding of the accuracy of the calculations for heavy actinides, it is instructive to compare electron structure factors for the HFS constants with those of lighter atoms, Dy, Ho, and Er. The situation is different for the HFS constants $A$ and $B$. The electron structure factor for the magnetic dipole constant $A$ is almost the same for heavy actinides and their lighter analogs; it varies within 3\%. The electron structure factors for the HFS constant $B$ are also similar, although the variation is larger. It goes from about 20\% for the Dy, Cf pair to 50\% for the Ho, Es pair. 
This justifies using lighter analogs of heavy actinides for the estimation of the uncertainty of the calculations. We assume 3\% uncertainty for the HFS constant $A$ of all considered atoms and 16\% uncertainty for the HFS constant $B$ (as the difference between theory and experiment for the ground state of Ho).
This latter assumption is rather conservative. The difference between theory and experiment for the HFS constant $B$ of the ground state of Dy is about 3\% and it is about 10\% for the ground state of Er~\cite{ErFm}.

This high level of accuracy is a bit surprising for atoms with open shells.
Therefore, it is instructive to see how dominating contributions are formed.
First, we note that according to numerical tests, configuration mixing gives a relatively small contribution to the HFS constants. About 90\% or more comes from leading configurations which is $4f^n6s^2$ for Dy and Ho and $5f^n7s^2$ for Cf and Es (n=10,11). In these configurations $s$-electrons form a closed shell and do not contribute to the HFS. Therefore, all contribution comes from $f$-electrons. It is well known that in the case of excited valence $f$-states (e.g, $4f$ state of Cs or $5f$ state of Fr) the HF value of the energy shift due to HFS operator $\langle 4f|\hat f| 4f\rangle$ is small and dominating contribution comes from the core polarisation correction $\langle 4f|\delta V_{\rm core}^f| 4f\rangle$ (see, Eq.~(\ref{e:de})). The situation is different in atoms considered in present work. The $f$ electron states are inside the core, localised at about the same distances as other states with the same principal quantum number, i.e. it is not even the outermost shell. For example, $\langle 4f|r| 4f\rangle < 1a_B$ for Dy, Ho and Er, while $\langle 4f|r| 4f\rangle \sim 20a_B$ for Cs.
Being inside the core $f$-states penetrate to short distances near the nucleus making a large value of the HF matrix element $\langle 4f|\hat f| 4f\rangle$.
In contrast, the core polarization correction $\langle 4f|\delta V_{\rm core}^f| 4f\rangle$ is small ($\sim$~1\%). In the end, zero-order matrix elements are large while core polarization and configuration mixing corrections are small. This is the key to the high accuracy of the results.

\begin{table*}
	\caption{\label{t:EsHFS}
		Hyperfine structure constants $A$ and $B$ (in MHz) of the ground state of Es. Nuclear spin $I$, nuclear magnetic moment $\mu(\mu_N)$, and
		nuclear electric quadrupole moment $Q(b)$ values for the isotopes of the $^{253}$Es are taken from Ref.~\cite{Stone1}, while $^{254}$Es and $^{255}$Es parameters are taken from Ref.~\cite{Raeder2022}.  $g_I=\mu/I$. The last column presents references for experimental data on $A$ and $B$.
		The values of $\mu$ and $Q$ obtained in this work are extracted from comparison of experimental and calculated HFS constants assuming 3\% uncertainty in calculation of $A$ and 16\% uncertainty in calculation of $B$.} 
	\begin{ruledtabular}
		\begin{tabular}{cc  cc c cc c cc}

			\multicolumn{1}{c}{Isotope}&&&
			\multicolumn{4}{c}{ This work}&
			
			\multicolumn{3}{c}{Experimental results.}\\
			\cline{4-7}
			\cline{8-10}
			\multicolumn{1}{c}{Nuclear Parameters}&
			\multicolumn{1}{c}{Conf.}&
			\multicolumn{1}{c}{Term}&
			
			\multicolumn{1}{c}{$A$}&
			\multicolumn{1}{c}{$B$}&
			\multicolumn{1}{c}{$\mu$}&
			\multicolumn{1}{c}{$Q$}&
			
			\multicolumn{1}{c}{$A$}&
			\multicolumn{1}{c}{$B$}&
			\multicolumn{1}{c}{Ref.}\\
			
			\hline

			\multicolumn{1}{c}	{\bf $ ^{253} $Es} \\
			
			$\mu$= 4.1(7), I= 7/2, Q= 6.7(8)	&	$5f^{11}7s^2$& $ ^{4} $I$ ^{o}_{15/2}$	 &798& -5481& 4.12(15)& 4.8(1.0)& 802(18)&  -3916(550)&\cite{Raeder2022}\\
			&&&&& 4.20(13)&5.3(8) &817.153(7) &  -4316.254(76)&\cite{Es1975}\\

			\hline
			\multicolumn{1}{c}	{\bf $ ^{254} $Es}\\
			
			$\mu$= 3.42(7), I= 7, Q= 9.6(1.2)	&$5f^{11}7s^2$& $ ^{4} $I$ ^{o}_{15/2}$	&333&-7853& 3.48(10)& 7.6(1.3) & 339(4)&  -6200(300)&\cite{Raeder2022}\\	 
			
			\hline
			\multicolumn{1}{c}	{\bf $ ^{255} $Es}\\
			
			$\mu$= 4.14(10), I= 7/2, Q= 5.1(1.7)	&$5f^{11}7s^2$& $ ^{4} $I$ ^{o}_{15/2}$ 	&806&-4172& 4.23(26)& 3.7(1.8)& 824(45) & -3001(1400)&\cite{Raeder2022}\\	
		\end{tabular}			
	\end{ruledtabular}
\end{table*}
	
Table \ref{t:EsHFS} shows the results and analysis of the HFS for three isotopes of Es ($^{253-255}$Es). This table serves two purposes. First, this is another confirmation of the accuracy of the calculations. However, to compare the calculations to the experiment we need to use nuclear moments, which are known to have pretty poor accuracy (see the table). For example, the uncertainty for the magnetic moment of the $^{253}$Es nucleus is 17\%. On the other hand, our estimated accuracy for the HFS constant $A$ is 3\%. This means that we can improve the accuracy of the nuclear moments by extracting them from a comparison of the experimental data with our calculations. The results are presented in the table. We see that real improvement is obtained for $\mu$($^{253})$Es only. For other nuclear moments, the uncertainties are similar but central points are shifted. New and old values are consistent when error bars are taken into account. 
	
\section{Conclusions}

Magnetic dipole and electric quadrupole HFS constants $A$ and $B$ were calculated for the ground states of heavy actinides Cf and Es.
Similar calculations were performed for the lighter analogs of these atoms, Dy and Ho. To establish the accuracy of the results, the comparison between theory and experiment was done for HFS constants, energy levels, $g$-factors and ionization potential, everywhere where the experimental data are available. 
We found the uncertainty of 3\% for the HFS constant $A$ and about 16\% uncertainty for the HFS constant $B$. 
Using the calculated HFS constants of those heavy elements considered, nuclear magnetic and electric quadrupole moments can be extracted from the measurement data.	
\begin{acknowledgments}

The authors are grateful to Sebastian Raeder for many stimulating discussions.	
This work was supported by the Australian Research Council under Grants No. DP190100974 and No. DP200100150. S.O.A. gratefully acknowledges the Islamic University of Madinah (Ministry of Education, Kingdom of Saudi Arabia) for funding his scholarship.
\end{acknowledgments}

\end{document}